\documentclass[]{aa}

\usepackage{txfonts}

\usepackage{graphicx}

\usepackage{natbib}
\bibpunct{(}{)}{;}{a}{}{,}
\newcommand\picsit{PICsIT}
\newcommand\isgri{ISGRI}
\newcommand\ipicsit{IBIS/PICsIT}
\newcommand\iisgri{IBIS/ISGRI}
\newcommand\spiacs{\mbox{SPI-ACS}}
\newcommand\kev{keV}
\newcommand\mev{MeV}
\newcommand\mygrb{\object{GRB\,030406}}
\newcommand\Nsigma[1]{#1$\sigma$}

\newcommand\peak{{\em peak}}
\newcommand\precursor{{\em precursor}}
\newcommand\tail{{\em tail}}

\begin{document}


\title{\mygrb\ -- an extremely hard burst\\
        outside of the INTEGRAL\thanks{INTEGRAL
        is an ESA project with instruments and science data center funded by the
        ESA members (especially the PI countries: Denmark, France, Germany,
        Italy, Switzerland, Spain), Czech Republic, and Poland, and with the
        participation of Russia and USA} field of view}\titlerunning{\mygrb}

\author{R.  Marcinkowski\inst{1,2} \and
         M.  Denis\inst{1}   \and
         T.  Bulik\inst{3,4}   \and
         P.  Goldoni\inst{5,6} \and
         Ph. Laurent\inst{5,6} \and
         A.  Rau\inst{7}
  }
\offprints{R. Marcinkowski, \email{radek@cbk.waw.pl}}
\authorrunning{R.Marcinkowski, M.Denis, T.Bulik }

\institute{Space Research Center, Bartycka 18a, Warsaw, Poland
             \and IPJ, 05-400 Swierk/Otwock, Poland
             \and Astronomical Observatory, Warsaw University, Aleje Ujazdowskie 4, 00478  Warsaw, Poland
             \and Nicolaus Copernicus Astronomical Center, Bartycka 18, 00716 Warsaw, Poland
             \and SAp CEA, 91911 Gif-sur-Yvette, France
             \and UMR 7164, 11 Place M. Berthelot, 75231 Paris, France
              \and Max-Planck Institute for extraterrestrial Physics, Giessenbachstrasse, 85748 Garching, Germany }

\date{Received date / Accepted date }

\abstract{Using  the IBIS Compton mode, the INTEGRAL satellite is able to 
detect and localize bright and hard 
GRBs, which happen outside 
of the nominal INTEGRAL field of view. We have developed a method of
analyzing such INTEGRAL data to obtain the burst location and spectra. 
We present the results for the case of \mygrb. 
The burst is localized with the Compton events, and the location is consistent 
with the previous Interplanetary Network position. A spectral analysis is 
possible by detailed modeling of the detector response for such a far 
off-axis source with an offset of
36.9\,$^\circ$. The average spectrum of the burst is extremely 
hard: the photon index above 400\,\kev\ is $-1.7$, with no evidence of a break 
up to 1.1\,\mev\ at a 90\% confidence level.
\keywords{gamma rays: burst}}

\maketitle

\section{Introduction}
The INTEGRAL satellite detects gamma-ray bursts (GRBs) in two different
ways: for a small number of events that fall in the field of view (FoV) of the
imager  IBIS \citep{2003A&A...411L.131U} and of the spectrometer SPI 
\citep{2003A&A...411L..63V},  INTEGRAL provides accurate positions  ($\sim$2 
arcmin) for rapid ground- and  space-based follow-up observations 
\citep{2003A&A...411L.291M}. A significantly larger number of GRBs occurs outside of the 
FoV of the two instruments.  These bursts  can be monitored  by the  SPI 
anti-coincidence system: \spiacs\ 
\citep{2003A&A...411L.299V,2005A&A...438.1175R}. Up  to now, the localization of 
the bursts that happened outside of the field of view was only possible with the 
aid of the 3$^\mathrm{rd}$ IPN \citep{1997trun.conf..491H}.

In this paper we show that for some of these bursts it is also possible to 
perform a more detailed  localization analysis using the Compton mode of IBIS, 
provided that the burst is sufficiently strong and spectrally hard.

We present the capabilities of the Compton mode using an example of \mygrb, 
a burst that was detected outside of the INTEGRAL field of view.
\mygrb\ has been detected by \spiacs\ on-board INTEGRAL as well as by Ulysses, 
Konus, and Mars Odyssey \citep{Hurley03}.
It was a long burst lasting $\sim 65$ seconds.
Ulysses reported the fluence of $1.3\times 10^{-5}$\,erg\,cm$^{-2}$ 
(25--100\,\kev). The observations by the three satellites yielded an estimate of 
the position centered on RA(2000) $= 19^h\,1^m\,43.0^s$, DEC(2000) $= 
-68^\circ\,4'\,39.4''$, with the \Nsigma{3}\ error-box of 77 square arcminutes.

In section \ref{sec:detection} we describe the Compton mode of IBIS INTEGRAL and 
demonstrate its localization capabilities. In section \ref{sec:spectral} we 
present the spectral analysis of \mygrb\ while section \ref{sec:discussion} 
contains the discussion.

\section{Observations and data analysis}
\label{sec:detection} 
The IBIS telescope is an imaging instrument on-board the INTEGRAL satellite with 
a coded mask \citep{2003A&A...411L.223G}. There are two detection layers in the 
detector plane \isgri\ \citep{2003A&A...411L.141L} and 
\picsit\ \citep{2003A&A...411L.149L}. \isgri\ is an array ($128\times128$) of 
pixels made of semiconductive CdTe, sensitive to photons between 15\,\kev\ and 
$\sim$1\,\mev. \isgri\ works in photon-by-photon mode. \picsit\ having the same 
detection area as \isgri, is an array of $64\times64$ CsI scintillators, 
sensitive between $\sim$170\,\kev\ and 15\,\mev\ located 94\,mm below \isgri. 
\picsit\ (usually) makes light curves of the whole detector by 
making histograms
of all events (so called Spectral Timing Mode). \isgri\ and \picsit\ can 
act as a Compton telescope, registering photons that are scattered in one and 
absorbed in the other detector. The coincidence time window is a parameter 
programmable on-board and it was set to $\sim 4$ microseconds at the time of 
\mygrb. The Compton mode is sensitive between 200\,\kev\ and $\sim$5\,\mev.

There is a 3 meter long collimator on top of the detector unit. The walls
of the collimator are made of lead and act as a shield to photons 
with energies up to $\approx 200$\,\kev. For the geometrical reasons the 
optical depth of the shield is smaller for photons arriving at large
angles. Thus hard photons from  off-axis sources can pass through the 
shield and reach the detector. In particular off axis GRBs may be detectable
in IBIS.

Thanks to the Compton mode data we were able to make successful, independent 
burst localization. Including information from the \isgri\ data we have performed burst
spectral analysis.

\begin{figure}
\includegraphics[width=0.9\columnwidth]{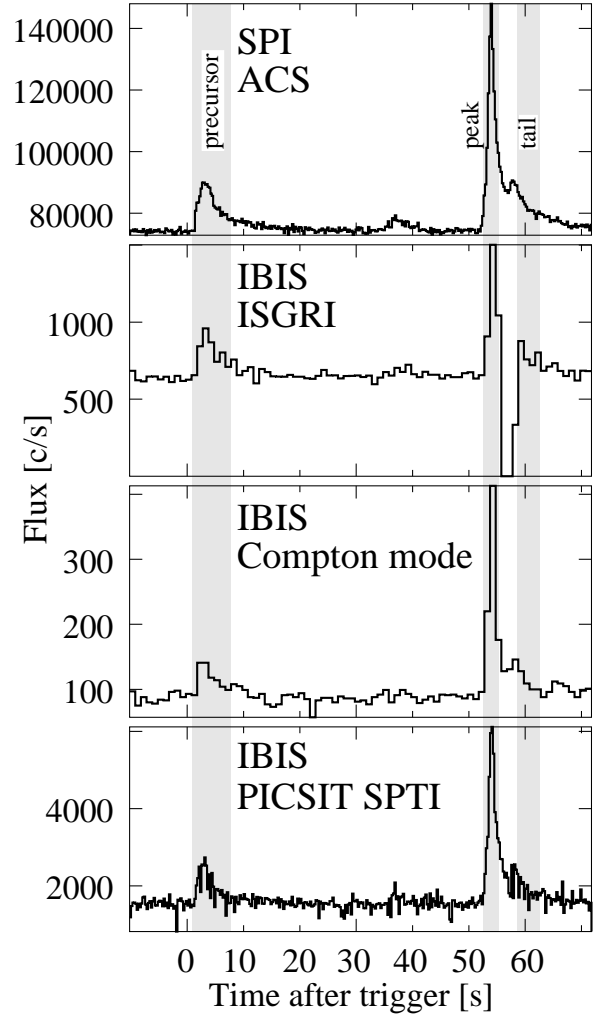}
\caption{Lightcurve of \mygrb\ seen by the INTEGRAL sub-systems: \spiacs, \iisgri, IBIS 
Compton and \ipicsit. The gray areas: \precursor, \peak\ and \tail\ correspond to the 
time regions chosen for spectral analysis. Time = 0 corresponds to the 
\spiacs\ trigger time at UTC 2003-04-06, 22:42:03.23.}
\label{lc}
\end{figure}

\subsection{Compton imaging and localization}
We have analyzed the Compton mode count rate during the \mygrb. 
The burst profile in the Compton mode follows clearly the profiles seen by 
\spiacs, \isgri\ and \picsit\ (Fig.~\ref{lc}). The gamma ray burst was detected 
in the Compton mode at the level of \Nsigma{30}\ reaching $\sim$400 
cts\,s$^{-1}$ in the second peak (average background amount to $\sim$100 
cts\,s$^{-1}$). The Compton mode telemetry did not suffer the data 
gaps in contrast to \isgri.

The Compton mode provides us with the following information:

\noindent -- energy deposits in \isgri\ ($E_\mathrm{I}$) and \picsit\
   ($E_\mathrm{P}$),

\noindent -- position of the detection in \isgri\ ($x_\mathrm{I},y_\mathrm{I}$)
  and  \picsit\ ($x_\mathrm{P},y_\mathrm{P}$),

\noindent -- timing of the event.

The two positions provide the information about the direction of the scattered 
photon. The two energies provide the Compton scatter angle $\theta_C$, which 
cosine, in case of the forward scatter, is given by:
$$ 
\cos \theta_\mathrm{C}  = 1 - 
\frac{m_\mathrm{e} c^2}{E_\mathrm{I}} + 
\frac{m_\mathrm{e} c^2}{E_\mathrm{I}+E_\mathrm{P}}
$$
where $m_\mathrm{e} c^2$ is the electron rest energy.
Given the angle $\theta_\mathrm{C}$ and the direction after scattering we obtain 
a ring in the sky which contains the possible directions of the primary photon.
In case of a point source all the rings cross at the location of the source.
For a given observation a collection of such rings provides a Compton map of the 
sky $S$.
In real observations the Compton map $S$ consists of the source signal and the 
background $B$.

In the case of \mygrb\ we selected the Compton events from the begin of the \peak\ up 
to the end of the \tail, which consist of a 9.75\,s time interval, see Fig.~\ref{lc}.
For 
these events a Compton map $S_{ij}$ of the sky was calculated using the method 
described above with the angular resolution of one degree. In order to determine 
the background we used the pre-burst data divided into 200 intervals lasting 
9.75\,s each. For each interval we calculated the all-sky Compton map and we 
determined the mean background $B_{ij}$ and its variance $V_{ij}$ in each 
pixel.

\begin{figure*}
\centering \includegraphics[width=17cm]{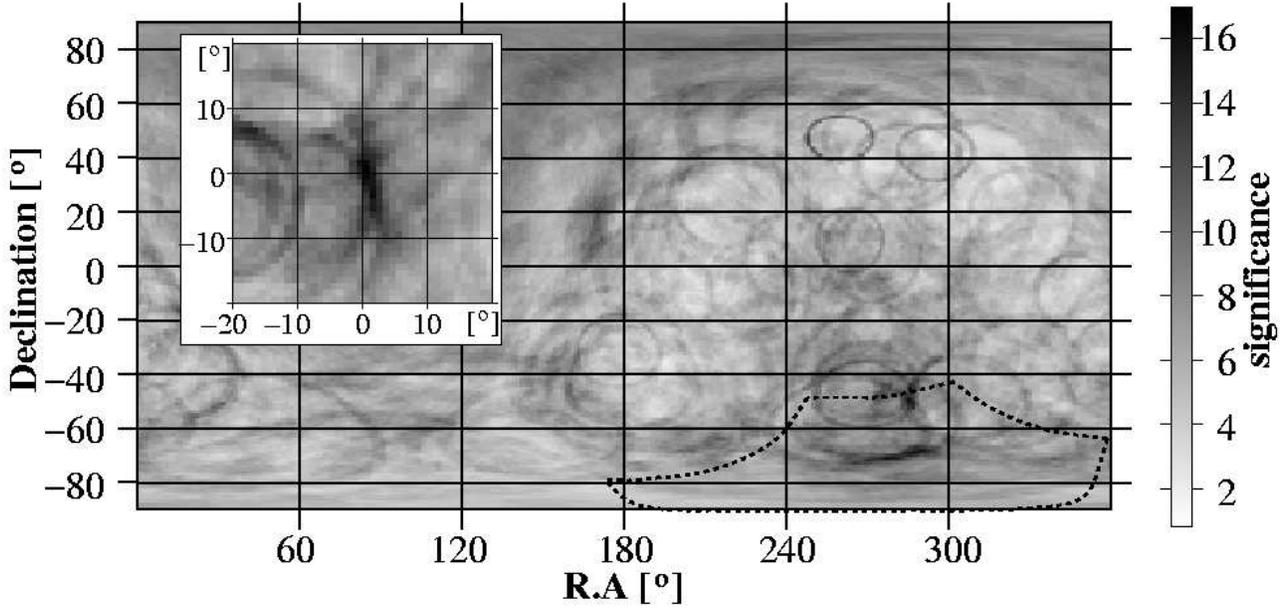} 
\caption{All-sky Compton map taken 
during the second peak of the \mygrb.  The inset shows the dashed area around 
the \mygrb\ centered on the IPN position.}
\label{local}
\end{figure*}

The map $\sigma_{ij}$ was obtained by dividing the difference between the 
Compton map $S_{ij}$ and the background $B_{ij}$ by the map of the square root 
of variance of the noise in each pixel $\sqrt{V_{ij}}$. We present the map of 
significance $\sigma_{ij}$ in Fig.~\ref{local}. The largest deviations from the 
noise are $\sim$20, and the position of the pixels with the strongest deviation 
is consistent with the IPN location.

In order to obtain an estimate of the accuracy of such positioning we used two 
approaches: the maximum likelihood and the Monte-Carlo. In the first approach we 
assumed that the probability density that the true position lies in a given 
pixel $i,j$ is proportional to the likelihood function:
$$
{\cal L}_{ij} \propto \exp\left( {\sigma_{ij}} \right).
$$
After normalizing such probability density we found the region containing 68\% 
of the probability corresponding to the $\sim 3.5^\circ$ radius.

\begin{figure}
\includegraphics[width=0.9\columnwidth]{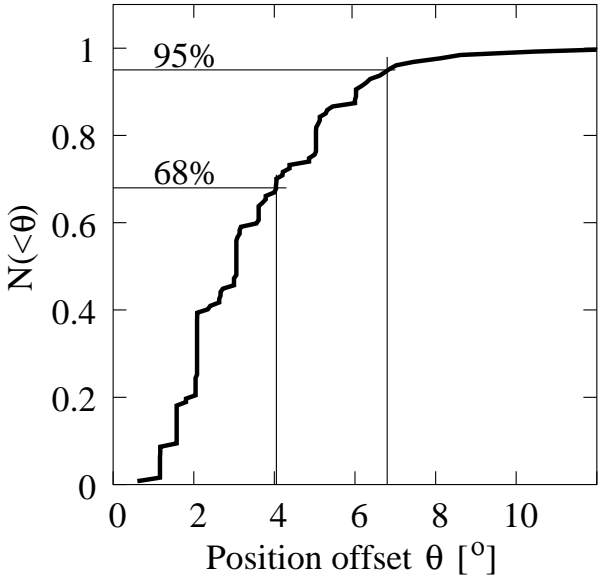}
\caption{The normalized cumulative distribution of the offset of best fit
   localizations of simulated bursts obtained in the Monte Carlo simulation: 68\%
   of the events are localized within  ~$\sim 4^\circ$ of the actual position.}
\label{simul}
\end{figure}

In the second approach we used the IBIS mass model \citep{2003A&A...411L.185L} 
to extract the signal from simulated GRB events inserted into stretches of data 
with experimental background. We then performed the identical data analysis 
procedure and found the localization of such simulated bursts. We have analyzed 
127 simulated bursts, and found the cumulative distribution of the distance 
between the actual burst position and the estimate obtained in our analysis. 
This distribution is presented in Fig.~\ref{simul}. The \Nsigma{1}\ region 
(containing $68$\% of the cases), corresponds to the accuracy of 
~$\sim$4$^\circ$, and the \Nsigma{2}\ is constrained to $6.5^\circ$. The two 
methods lead to consistent 
results.

\subsection{Spectral analysis}
\label{sec:spectral}
In addition to the Compton mode 
\mygrb\ has been also detected by \isgri\ and \picsit. There are only two energy 
channels in the \picsit\ data stream therefore we have decided to model together
the Compton and \isgri\ data, neglecting the \picsit. 
\begin{figure*}
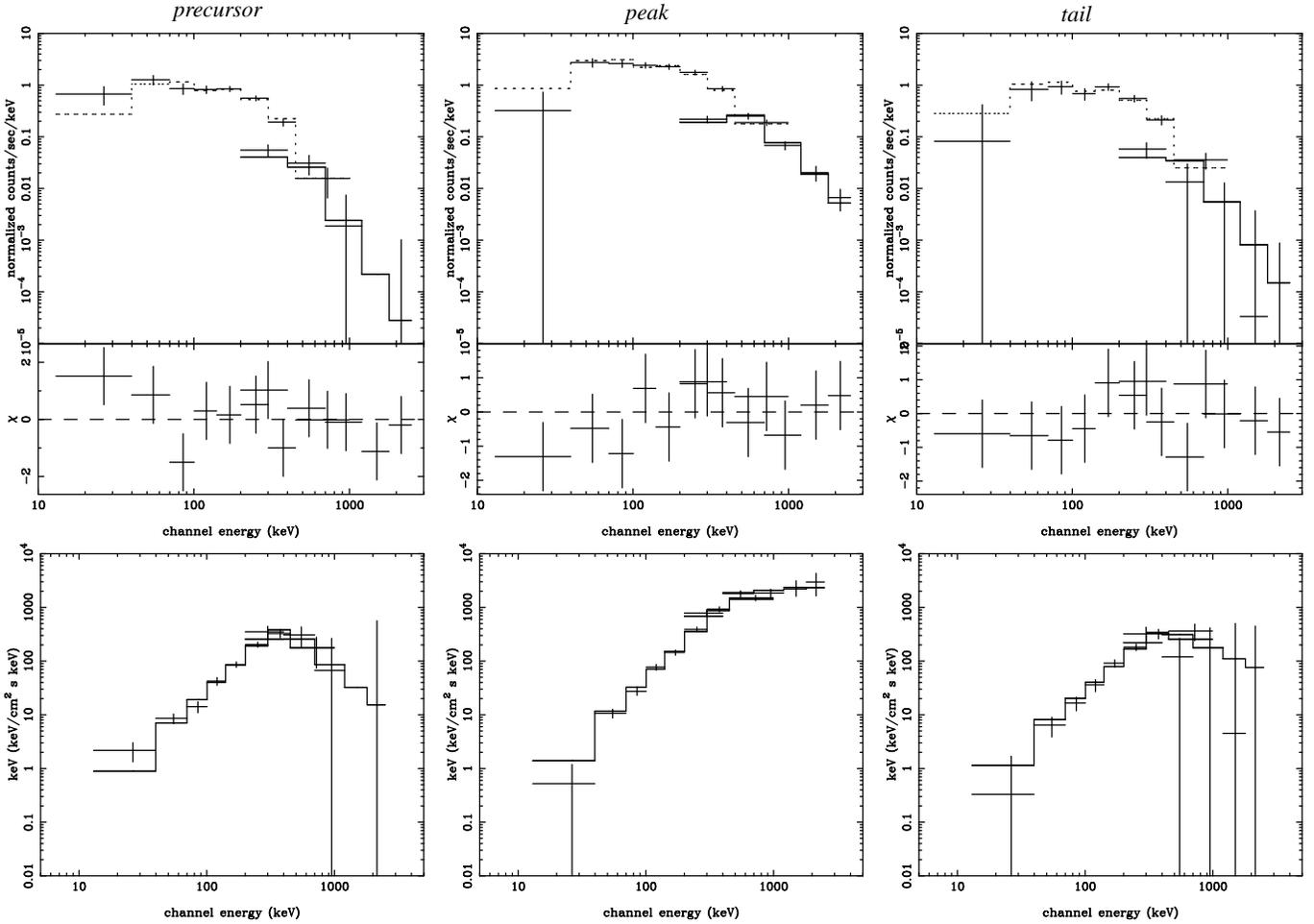

   \begin{minipage}{0.33\hsize} \centering \precursor \end{minipage}%
   \begin{minipage}{0.33\hsize} \centering \peak      \end{minipage}%
   \begin{minipage}{0.33\hsize} \centering \tail      \end{minipage} \\
   \resizebox{\hsize}{!}{\includegraphics[angle=0]{4811fig4.eps}\hspace{0.7cm}\includegraphics[angle=0]{4811fig8.eps}\hspace{0.7cm}\includegraphics[angle=0]{4811fig6.eps}}
   \resizebox{\hsize}{!}{\includegraphics[angle=0]{4811fig5.eps}\hspace{0.7cm}\includegraphics[angle=0]{4811fig9.eps}\hspace{0.7cm}\includegraphics[angle=0]{4811fig7.eps}}
   \caption{\mygrb\ spectrum fits to the combined \isgri\ and Compton mode data.
   Upper panels show count spectra, while lower corresponding $\nu \mathrm{F}_\nu$ spectra.
   Broken power law model was fitted in each case: to the \precursor\ data (left),
   to the \peak\ part data (center) and to the \tail\ (right). See text for details}
   \label{spec}
\end{figure*}

We have divided the data into three time intervals
marked in gray in Figure~\ref{lc} as: \precursor, \peak\ and \tail.
The division of the main peak of the burst 
into the \peak\ and \tail\ was a
consequence of the \isgri\ data loss during the burst
due to telemetry gap.

We first ran Monte Carlo simulations using the mass model 
of the INTEGRAL spacecraft to generate the detector response matrices (DRM) for 
a source located at the position of the burst. The DRM matrices were obtained 
for the \isgri\ and Compton sub-systems. For all of them we have taken into
account the influence of the IBIS VETO system, 
the energy thresholds in each \isgri\ pixel and \picsit\ crystal, the 
\isgri\ noisy pixels, the failed 
\isgri\ pixels and \picsit\ crystals as well as 
the  \isgri\ and \picsit\ energy resolution.


The Compton data were binned into six 
energy channels covering the range from 200 to 2500\,\kev.
The \isgri\ data spans from 24 to 450\,\kev\, 
which partially overlaps the Compton energy range and were binned into seven channels. 
The analysis of the pre-burst data with the correction for  the varying 
dead time effect and exposures provided an estimate of the background. This 
background was subtracted from the total count spectrum in each channel.

We used the standard XSPEC 11.2 \citep{1996ASPC..101...17A} to fit the data.
For each time region we tried to fit several models: single power law, 
broken power law,  and a non physical model 
of single  black-body.  The single power law model never fitted the combined \isgri\ and Compton
data.
The broken power law fits the data very well in all analyzed regions:
\precursor, \peak\ and \tail. In additional fit we have allowed for 
different normalization of \isgri\ and Compton components for \peak\ 
region, taking to account two facts related with dead time influence:

\noindent -- during \precursor\ and \tail\ the count rate
grows with factor about 1.5 for \isgri\ and Compton mode, while during
\peak, these factors reach 2.5 for \isgri\ and 4 for Compton mode,

\noindent -- the width of the burst peak, $\approx$1s,
is much shorter than the INTEGRAL dead time effect calculation bin, equal
8s.

\noindent Except meaningful decreasing of the reduced $\chi^2$ we have not
find significant evidence of changing fitted parameters.
Results of all fits are presented in Table~\ref{tab:fit}.
The broken power law fits and the residuals are presented in the top panels 
of  Figure~\ref{spec}.

We note that the precursor and the tail region can also be fitted with
  black body spectra with temperatures $\approx 100$\,\kev. This is due to the 
fact that at high energies the count rates fall steeply.
The lower panels of Figure~\ref{spec} show the deconvolved $\nu F_\nu$ spectra
in each time interval. The \precursor\ and \tail\ spectra have peak power
around 300-500\,\kev, while the $\nu F_\nu$ spectrum in  the \peak\ seems to rise 
all the way into high energies.

\begin{figure}
\begin{center}
   \resizebox{\hsize}{!}{\includegraphics{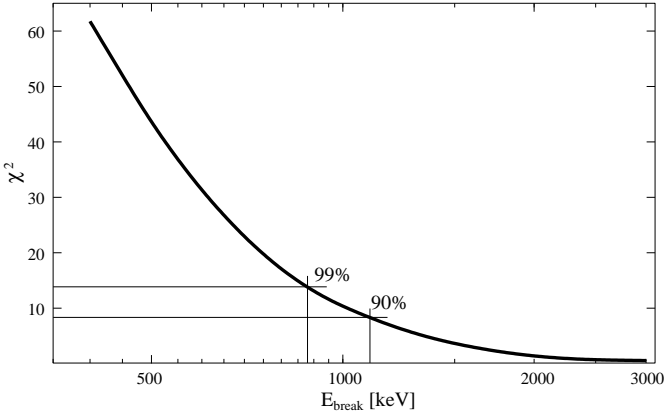}}
\end{center}
\caption{Confidence limits on $E_\mathrm{cutoff}$: the solid line 
represents the $\chi^2$ as a function of $E_\mathrm{cutoff}$ using the 
model of equation~(\ref{eq:chi}). We present the  $\chi^2$  values 
corresponding to the 90\% and 99\% confidence levels.}
  \label{chi}
\end{figure}

The high energy index of $\beta=-1.7 \pm 0.3$ for \peak\ may imply 
the absence of a peak in the source power spectrum $\nu F_\nu$ in the range covered by our 
data, i.e. up to 2.5\,\mev. In order to estimate the lower limit on such a break 
compatible with the data we conducted the following analysis. We fit the Compton 
data    with a broken power law with fixed high energy index:
\begin{equation}
f(E) = N 
\left\{
   \begin{array}{cl}
            E^\beta & E < E_\mathrm{break} \\
            E_\mathrm{break}^{\beta+4} E^{-4}  & E > E_\mathrm{break}
   \end{array}
\right.
\label{eq:chi} 
\end{equation}
In this initial fit the  break energy was constant
  $E_\mathrm{break}$=3\,000\,\kev.
This value was chosen high enough not to influence the power law part of the fit.
  For this fit  we obtained  $\chi^2=0.55$ with 4 degrees of freedom.
We used the standard model parameter estimation \citep{1976ApJ...208..177L}
to calculate the  values of $\chi^2$ corresponding to  90\% 
and 99\% confidence regions, which are respectively $\Delta \chi^2=8.3$
and $\Delta \chi^2=13.8$.
Then keeping all  the parameters but $E_\mathrm{break}$ constant
we looked for the lowest
value of $E_\mathrm{break}$ for which $\chi^2$ was equal to the above
  values. We present the results in Figure~\ref{chi}.
We obtained $E_\mathrm{break}>$1\,110\,\kev\ at  90\%  confidence level
and $E_\mathrm{break}>$880\,\kev\ at 99\% confidence level.

%
%

\begin{table*}
\caption{Results of the spectrum fits for three time ranges of the \mygrb. All the errors are at the $1\sigma $ level.}
\label{tab:fit}
     \centering
     \begin{tabular}{c  c  r r r c  r c}
     \hline
     \hline
\bf Part of & \bf duration & \multicolumn{4}{c}{ \bf broken power law fit }              &  \multicolumn{2}{c}{\mbox{\hspace{1cm}} \bf blackbody fit \mbox{\hspace{1cm}}}\\
      \bf the GRB & [s]          & $\alpha$              & $\beta$             & $E_{break}$ [keV] & $\chi^2/$dof & kT [keV]        &  $\chi^2/$dof \\
     \hline
      precursor   & 7.3          &  $0.0^{+0.3}_{-0.3}$  & $9.0^{+1}_{-6}$     & $490^{+40}_{-180}$&    0.96      & $106^{+12}_{-13}$ & 0.82          \\

&&&&&&&\\
  peak            & 2.81         &  $-1.5^{+0.7}_{-1.0}$ & $1.7^{+0.4}_{-0.3}$ & $390^{+60}_{-50}$ &    1.19      & \multicolumn{2}{c}{\bf ---}     \\

  peak$^\dagger$  &              &  $-2.5^{+1.5}_{-0.5}$ & $1.7^{+0.2}_{-0.2}$ & $300^{+30}_{-30}$ &    0.49      & \multicolumn{2}{c}{\bf ---}     \\
&&&&&&&\\
      tail        & 4.3          &  $-0.8^{+0.7}_{-2.2}$ & $2.8^{+1.2}_{-0.6}$ & $270^{+70}_{-50}$ &    0.56      & $140^{+20}_{-20}$ & 0.56          \\

     \hline
     \end{tabular}

     NOTE: $^\dagger$ data fitted with variable  normalization for both sub-instruments. See text.
\end{table*}

\section{Discussion}
\label{sec:discussion}
We have shown that the IBIS Compton mode is able to detect and localize GRBs 
outside of the field of view of the INTEGRAL telescopes. We have found that the position of 
\mygrb\ obtained using the Compton mode technique is consistent with the IPN 
localization. The position accuracy has been estimated using the maximum 
likelihood and the Monte Carlo method. The results of both methods are consistent 
and the \Nsigma{1}\ errors are $\sim 4^\circ$.

  Given the position of the burst 
we computed the DRM for a source outside of the field of view of the satellite 
at this particular location. Using the  DRM we analyzed the the Compton data, 
along with the \isgri\ detection and fitted the  spectrum with a broken power 
law in the range from 50\,\kev\ to 3\,\mev.

The \peak\ spectrum is very hard; in the 
$\nu F_\nu$ the low energy index below 400\,\kev\ rises with the index $\approx 
+3.5$ and above this energy it is still positive $\approx +0.3$. We note that 
this spectrum is an average over the time interval of 2.8s around the maximum, thus 
the spectrum of the maximum might have   even been harder 
since the burst spectra usually 
evolve from hard in the peaks to soft in their tails. We do not see any evidence 
that  the $\nu F_\nu$ spectrum peaks below 1.1\,\mev. Thus, there is a hint of existence of
   bursts with the 
peak of $\nu F_\nu$ spectrum above the distribution shown by 
\citet{1995ApJ...454..597M} which truncates at $\sim$1\,\mev\ for the bright 
BATSE bursts.

Such hard spectra have already been seen by BATSE -- 
\citet{2000ApJS..126...19P} present the distribution of the low and high energy 
spectral indexes for a sample of 156 bright bursts.
   Both, the low and the high energy spectral indexes of \mygrb\ lie in 
the upper tails of the distributions of the spectral indexes presented by 
\citet{2000ApJS..126...19P}. It is therefore clear that the spectrum of this 
particular burst is in clear contradiction with the synchrotron model of GRBs 
\citep{1994ApJ...432L.107K,1995Ap&SS.231..181T}, which predicts that there is a 
strict upper limit on the low energy spectral index of $-{2\over 3}$ 
\citep{1998ApJ...506L..23P}. The low energy spectral slope is consistent within 
the error bars with the jitter synchrotron model of GRBs 
\citep{2000ApJ...540..704M}. We note that a detailed study of time resolved 
spectra \citep{2003A&A...406..879G} showed that the low energy spectral slopes 
are large $\sim 0.5$ -- $1$ at the rising parts of several bursts.

The value of $E_p$ of \mygrb\ -- the peak of the 
$\nu F_\nu$ spectrum lies above 1.1\,\mev. It should be noted that this is on the 
high end of the Amati and Ghirlanda relations \citep{2002A&A...390...81A,2004ApJ...616..331G}.
  If this relation 
holds and we assume that the peak energy is larger than the $90$\% lower limit
of $1100$\,\kev,
then the isotropic radiated energy for \mygrb\ is larger than $(6-20)\cdot 
10^{53}$\,erg, where the uncertainty stems from the inaccuracy of the Amati relation.
Hard bursts are therefore very energetic and may 
come from a population residing at high redshifts.

In 
summary, we confirm  that the INTEGRAL in the Compton mode can detect hard GRBs.
  The localization accuracy is a few degrees. The GRB spectra can  be studied in 
the range from $\sim$200\,\kev\ to $\sim$3\,\mev\ depending on the location of 
the burst in the instrument coordinates. Further study of such hard bursts 
detectable in the INTEGRAL Compton mode may yield potential candidates of high 
redshift burst and also test the limits of validity if the Amati relation. Our 
rough estimations of the Compton mode capabilities give 2-3 detections of such 
bursts per year. They are consistent with BATSE results and seem to be confirmed 
by the other detections (Marcinkowski et al. in preparation). 
Finally, we note that   analysis 
of Compton data in principle allows to determine the polarization. A study of 
of the polarization sensitivity of the IBIS Compton mode is currently under way
(Laurent et al. in preparation).

\begin{acknowledgements}
This research 
was supported by the KBN grants 2\,P03D\,001\,25, 1\,P03D\,003\,27 and 
PBZ-KBN-054/P03/2001.
\end{acknowledgements}

\end{document}